\documentclass[journal,onecolumn,12pt]{IEEEtran}
\IEEEoverridecommandlockouts
\usepackage{amsmath, amssymb, stmaryrd} 
\usepackage{graphicx}
\usepackage{amsmath}
\usepackage{color}
\usepackage{hyperref}
\usepackage{multirow}
\usepackage{fancyhdr}
\usepackage[normal,bf,up]{caption2}
\usepackage{subfigure}
\usepackage{amsthm}
\usepackage[section]{placeins}
\usepackage{comment}
\usepackage{enumitem}
\usepackage{tabularx}
\usepackage{graphicx}

\usepackage{algorithm}
\usepackage{algpseudocode}

\usepackage{amssymb}
\usepackage{setspace}
\usepackage{mathtools, nccmath}

\begin{document}
\parindent=5mm

\DeclarePairedDelimiter{\nint}\lfloor\rceil
\newcommand{\beq}{\begin{equation}}
\newcommand{\eeq}{\end{equation}}
\newcommand{\Hb}{\mathbf {H}}
\newcommand{\F}{\mathbf {F}}
\newcommand{\B}{\mathbf {B}}
\newcommand{\I}{\mathbf {I}}
\newcommand{\p}{\boldsymbol {p}}
\newcommand{\bv}{\boldsymbol {b}}
\newcommand{\matr}[1]{{\mathbf{#1}}}
\renewcommand{\vec}[1]{{\boldsymbol{#1}}}
\newcommand{\x}{\boldsymbol {x}}
\newcommand{\n}{\boldsymbol {n}}
\newcommand{\s}{\boldsymbol {s}}
\newcommand{\w}{\boldsymbol {w}}
\newcommand{\vv}{\boldsymbol {v}}
\newcommand{\cx}{\boldsymbol {c}}
\newcommand{\rx}{\boldsymbol {r}}
\newcommand{\y}{\boldsymbol {y}}
\newcommand{\z}{\boldsymbol {z}}
\newcommand{\R}{\mathbf {R}}
\newcommand{\A}{\mathbf {A}}
\newcommand{\mL}{\mathcal{L}}
\newcommand{\raw}{\rightarrow}

\title{ Wireless-Powered Communication Assisted by Two-Way Relay with Interference Alignment Underlaying Cognitive Radio Network}
\author{
      Iman Pazouki \texttt{iman.pazouki@dal.ca}\\
  \and Roshanak Soltani \texttt{roshanak.soltani@dal.ca}\\
  \and Mohammad Lari \texttt{m\_lari@semnan.ac.ir}
}

\maketitle

\begin{abstract}
This study investigates the outage performance {\color{black}of} an underlaying wireless-powered secondary system that reuses the primary users’ (PU) spectrum in a multiple-input multiple-output (MIMO) cognitive radio (CR) network. {\color{black} Each secondary user (SU)  harvests energy and receives information simultaneously by applying power splitting (PS) protocol.} The communication between SUs is aided by a two-way (TW) decode and forward (DF) relay. We formulate a problem to design the PS ratios at SUs, the power control factor at the secondary relay, and beamforming matrices at all nodes to minimize the secondary network’s outage probability. {\color{black}To address} this problem, we propose a two-step solution. The first step establishes closed-form expressions for the PS ratios at each SU and secondary relay’s power control factor. Furthermore, in the second step,  interference alignment (IA) is used to design proper precoding and decoding matrices for managing the interference between secondary and primary networks. We choose IA matrices based on the minimum mean square error (MMSE) iterative algorithm. {\color{black}The simulation results demonstrate a significant decrease in the outage probability for the proposed scheme compared to the benchmark schemes, with an average reduction of more than two orders of magnitude achieved.}
\end{abstract}


\section{Introduction}\label{section.intro}
    %CR
Cognitive radio (CR) technology is among the crucial ways to achieve better utilization of the spectrum in wireless communication networks \cite{ElTanab2016}. In CR networks, the unlicensed users or secondary users (SUs) reuse the licensed users or primary users' (PUs) spectrum. There are mainly three approaches for SUs to use the licensed spectrum: interweave, overlay, and underlay \cite{Arzykulov2018}. SUs can use the part of the spectrum which is unused by PUs in the interweave method \cite{Awin2019}. In overlay mode,  SUs can have access to the spectrum by cooperating with PUs \cite{prathima2020performance}. Finally, the underlay technique, which is the scope of this study, allows SUs and PUs to communicate simultaneously and in the same frequency band \cite{ElTanab2016}. There are two tiers within the underlay CR network: the primary network tier and the secondary network tier. One of the most challenging issues related to the underlay CR is the inevitable interference between the signals of mentioned tiers \cite{amodu2021outage}. If interference is not appropriately managed, the consequences have adverse effects on both tiers' performance. Hence, instead of being useful, CR can disturb the whole communication in a system. Consequently, many studies attempted to diminish this negative effect on PUs by imposing a power constraint on SUs, namely interference temperature constraint (ITC). However, by applying a powerful interference mitigation method, we can eliminate the interference signals not just at PUs but at SUs, and guarantee an acceptable performance for both network tiers.

%IA
One of the well-known methods to mitigate the interference in a system is interference alignment (IA)\cite{Arzykulov2018,Zhao2016}. In the IA algorithm, to be concise, transmitters apply proper precoding matrices to their signals so that the interference signals at each receiver will be aligned in an independent subspace from their desired signals. Then, receivers eliminate interference signals by applying a decoding matrix which is placed in the orthogonal subspace. IA has been applied on many networks such as full-duplex networks \cite{Aquilina2017,wang2019interference}, wireless powered relay networks \cite{Chu2018}, and device-to-device networks \cite{wang2021joint}. In particular, using IA instead of imposing ITC in an underlay CR network and its advantages has been studied in \cite{Arzykulov2018}. Furthermore, \cite{namdar2022iterative} {\color{black}proposed} a novel approach for applying IA on a CR network by detecting and utilizing the unused degrees of freedom in the primary network to align interferences caused by secondary network users, enabling efficient secondary transmission. Moreover, in \cite{he2016}, a diversity-based IA aiming for coping with decreasing the signal-to-interference-plus-noise ratio (SINR) at PUs {\color{black} was} proposed.

%EH
In the not too far future, the increasing use of wireless communication devices, together with the slow progress in elevating batteries' capacity and the importance of mobility, will lead us towards either reducing the power consumption using methods such as antenna selection \cite{lari2020multi}, or utilizing wireless-powered devices \cite{Padhy2021}. Wireless energy harvesting (WEH) enables each node to provide its needed power from received signals, so it reduces the need for wire-charging considerably. In this context, energy can be sent to a wireless-powered node in the form of an energy signal separate from information signals \cite{lari2019transmission}. \cite{wu2018transceiver} {\color{black} suggested} a transceiver design for IA based CR networks in which there exists 2 groups of SUs: energy harvesters and data transmitters. Also, PUs and SUs apply IA to either partially or fully cope with their received interference signals. Moreover, radio frequency (RF) signals add a new aspect to WEH, namely simultaneous wireless information and power transfer (SWIPT) \cite{Shi2019,zahedi2017simultaneous}. SWIPT enables each node to transmit data and power concurrently through the RF signals, so the wireless-powered receiver can jointly provide energy and information from its received signals. So far, studies have shown that SWIPT is more practical for low-power devices within the small distances; otherwise, the path-loss effect might reduce the signals' power to a level that they can no longer serve the dual purpose. On the bright side, the increasing number of users and the tendency towards green communication direct wireless communication to the inevitable decrease of spaces between users and less power consuming, respectively, providing a favorable setting for SWIPT. There are two main protocols for implementing SWIPT in wireless systems: power splitting (PS) \cite{wang2020optimal}, and time switching (TS)\cite{lee2021robust}. PS outperforms TS in many cases since it makes a more efficient trade-off between the harvested power and transmission rate \cite{hossain2019survey}. In the PS protocol, the receiver divides its received signal into two parts with a fraction named the PS ratio, then sends one part to the information processing (IP) unit and the other one to the energy harvesting unit for providing power.

%EH IA
Applying WEH on interference channels allows each node to harvest the power of received interference signals. As a result, as long as we can manage the interferences well enough to control their disruption in information processing, interferences are valuable power resources \cite{Chu2018,Zhao2017,soltani2021performance,soltani2020wireless}. As we mentioned earlier, IA is an essential method for managing interferences, the effectiveness of which in WEH networks has studied comprehensively in \cite{Zhao2017,Zhao2018}. By equipping SWIPT users with IA technique, \cite{Xu2021} {formulated} an optimization problem to maximize the received SINR at each receiver in a MIMO multi-user system. In \cite{Chu2018} the design of IA beamforming matrices in wireless powered one-way relays together with choosing the optimal PS ratios { was} investigated. The authors in \cite{Arzykulov2018} {applied} a one-way DF wireless-powered relay for assisting the transmission between two SUs in a cognitive relay network. They {investigated} the performance of both PS and TS protocols for harvesting power at the secondary relay. Also, to manage interferences, they {used} IA at primary and secondary users. 

%relying
For transmitting information efficiently, specifically between low-power devices, relaying has been proved to be a practical approach \cite{Shi2019}. Applying relays increases the system's throughput considerably and extends the network coverage. Also, relaying reduces the power consumption of battery-limited users. All mentioned qualities makes relaying a reasonable match for SWIPT.  Moreover, in CR networks, using relays for assisting the communication within the secondary network can notably decrease the interference power caused by SUs at PU receivers \cite{ni2017}. Compared to one-way relaying, using two-way (TW) relays is more beneficial for improving the spectral efficiency since the information exchange between two nodes can occur in three time-slot or frequency bands instead of four. With three-step TW relays, at the first time-slot, one source sends data to relay. The second source sends information to the relay at the second time-slot, and finally, at the third time-slot, the relay forwards the information signals to both sources. Each source is aware of its transmitted signal and can remove it from its received signals to achieve the desired information. Furthermore, in TW decode and forward (DF) relays by controlling the power control ratio at the relay, we can divide the relay's transmit power between nodes to result in the system's performance growth. The performance of TW relay in the CR network has been investigated in \cite{van2018} with a single antenna setting. This study {explored} the probability of interference from SUs for PUs and the power control coefficient for the secondary network. Also, it {probed} the outage probability of the secondary network for evaluating its performance.  Furthermore, in \cite{li2017} using a multi-antenna TW relay in a CR network with the goal of minimizing the outage probability of the secondary network {was} explored. Considering a multiple-relay scenario, the authors proposed the proper schemes for relay selection and power allocation. {\cite{duy2021outage} proposed a digital network coding protocol in underlay cognitive radio, utilizing transmit antenna selection, selection combining, and interference cancellation to enhance secondary network performance.}

%relay & EH
As we mentioned earlier, combining SWIPT with relaying has several advantages, so it has been the subject of many studies aiming for performance improvement. For instance, \cite{Tian2020} {investigated} a multi-relay system in which SWIPT users provide their needed power by harvesting the received signals transmitted by two-way relays. In \cite{Nguyen2017}, the outage performance of a TW wireless powered amplify and forward (AF) relay in three scenarios (TW communication with 2,3 and 4 time-slots) has been investigated. It also {explored} a scenario in which the relay does not harvest the wireless power's energy for providing power to compare. The authors in \cite{zeng2018} {investigated} the outage probability of the SUs in the CR network with the assistance of a TW wireless powered relay, in which all nodes {were} equipped with a single antenna. In order to control the interference at PUs, the transmit power of SUs {was} restricted to a threshold value, which means SUs cannot communicate data unless their power is lower than a specific value. In \cite{Shi2019} the outage performance of a system consisting of two single antenna sources communicating over a DF wireless powered TW relay {was} investigated. Moreover, it {suggested} a dynamic scheme for choosing both PS and power control ratios at the relay for minimizing the outage probability. {\cite{tung2021performance} proposed a Two-Way Cognitive Relay Network (TWCRN) utilizing underlay mode, self-powered relay, and TS/PS methods, evaluating performance under hardware impairments and co-channel interference, with derived closed-form expressions for outage probability, throughput, energy efficiency, and optimal relay position. } {\cite{salari2023maximizing} presented an energy harvesting cooperative cognitive radio network (CCRN) framework, where the relays in the secondary network periodically switch between energy harvesting and information transmission tasks to achieve spectral and energy efficiency.} Unlike many studies in applying WEH on relay networks that choose the relay node to be wireless powered, in \cite{W.Wang2017}, multi-antenna relay {was} connected to a grid, and the single antenna sources {harvested} power from the relay transmitted signals with monotonous PS ratios. {Further in this context, in \cite{tang2021spectrum} the secondary transmitter harvests energy from primary users' RF signals, switches between silent and data transmission modes, and optimizes sub-slot switching and power allocation coefficients for enhanced spectrum efficiency and energy efficiency.}

    \subsection*{Our Contribution}
In this paper, we attempt to improve the performance of two SUs communicating over a TW DF relay in a multiple-input multiple-output (MIMO) CR network in the presence of PUs and their dedicated relay. {We make the assumption that the direct link between the SUs is negligible due to the significant path-loss and shadowing effects and can therefore be disregarded.} Unlike many studies in relaying areas that apply WEH at the relay, in this study, the SUs (the sources, not the relay) are wireless-powered. This scenario is closer to the real case since users are usually equipped with restricted-power batteries. Moreover, to eliminate the interference signals, we apply IA on each node in the underlay CR network instead of imposing ITC on SUs. To our best knowledge, this is the first study to combine wireless-powered transmission between SUs in a CR network with TW relaying and IA to achieve a convenient performance for future low-power communications. We formulate the outage probability minimization problem. Then, we present our solution method in two parts. In the first part, we drive a closed-form solution for achieving the optimized PS ratios at both wireless-powered SUs separately to make the best trade-off between energy harvesting and information transmission. In this part, we also propose a closed-form solution for achieving the power control ratio at the secondary relay to minimize the outage probability of SUs. In the second part, we eliminate interference signals by applying IA at all nodes using the minimum mean square error (MMSE) iterative method, {which is considered to be one of the best IA approaches \cite{soltani2021performance}.}

    The major contribution of this paper is abridged as follows.

    \begin{itemize}
      \item We formulate the outage probability minimization problem for the secondary network, and we propose the following 2-step solution to solve this problem.
      \item In the first step we derive a closed-form solution for achieving the optimized PS ratios at each SU separately to make the best trade-off between energy harvesting and information transmission. We also derive a closed-form solution for allocating the relay's transmit power to each SU aiming for minimizing the secondary network's outage probability. 
      \item In the second step, we design the IA beamforming matrices based on MMSE iterative algorithm in order to mitigate interference signals at each node in the network. As a result, by eliminating the received interference signals from PUs at each wireless-powered SU's IP unit, we assure that interference signals are almost completely beneficial for them since they not only do not disturb the information processing but also provide SU's needed power.
      \item Finally, with the simulations, we investigate the outage performance of our proposed scheme based on different parameters such as the secondary relay's transmit power, distance, threshold SNR, and antenna number. Simulation results show the superiority of the proposed scheme over the benchmark schemes in different scenarios.
    \end{itemize}

     The remainder of this paper is organized as follows. In Section \ref{section.model} the system model is presented. We provide the optimization problem for minimizing outage probability of the secondary network in Section \ref{section.solution} and then the proposed two-part solution is explained. Section \ref{section.results} exposes the outage performance results through simulations. Finally, Section \ref{section.conclusion} concludes this study.

\emph{Notation:} $(.)^H$, $\|.\|$, $rank(.)$, $\mathbf{I_l}$, $E\{.\}$, $Tr(.)$, $\mathbb{C}^{M\times N}$, and $\mathbb{CN}(a,b)$, denote the Hermitian transpose, $l_2$-norm, rank, the  identity matrix $l\times l$, the expectation operator, trace, the space of complex $M\times N$ matrices, and the distribution of a circularly symmetric complex Gaussian random vector with mean ${a}$ and covariance matrix ${b}$, respectively. {Furthermore, we show variables with italic, vectors with non-italic bold lower-case, and matrices with non-italic bold upper-case letters.}

\section{System Model}\label{section.model}
\begin{figure}[t]
  \centering
  \includegraphics[scale=0.75]{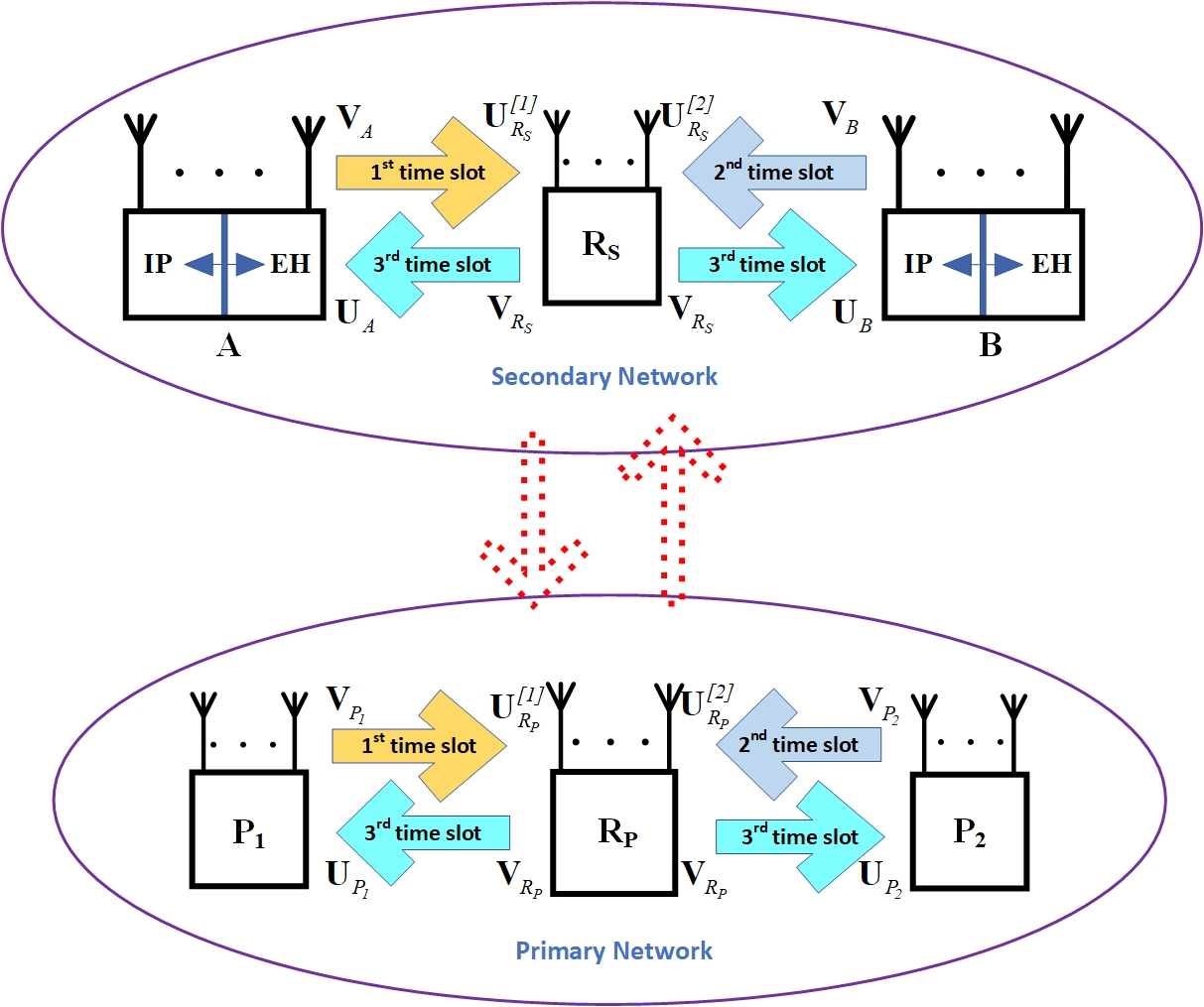}
  \caption{CR Network consisting of two wireless powered SUs communicating over a secondary relay, two PUs and their dedicated relay. Dashed-line arrows present the interference channel between secondary and primary network. }\label{fig:1}
\end{figure}

According to Figure \ref{fig:1}, we consider the performance of two wireless-powered SUs $(A,B)$  communicating over a TW DF relay $(R_S)$  in three time-slots. At the first time-slot, $A$ transmits data to the relay. The second time-slot is when $B$  communicates information to the relay. Finally, at the third time-slot, the relay forwards the decoded and reencoded signals to SUs. The strong path-loss effect attenuates the direct link between $A$ and $B$, and consequently, this link can be ignored. Besides the secondary network, there are two PUs in the system $(P_1, P_2)$ which intend to communicate over a TW DF relay $(R_P)$. Since the communication of SUs and PUs is simultaneous and in the same frequency band, their signals interfere with each other. We assume a MIMO setting in which each node is equipped with $N_k$ antennas for $k \in \{A,B,R_S,P_1,P_2,R_P\}$. In the secondary network, for $i \in \{A,B\}$, Rayleigh channels between $i-R_S$ are denoted with $\mathbf{H}_{{R_S}i} \in \mathbb{C}^{{N_{R_S}}\times {N_i}}$, and channels between $R_S-i$ are described with $\mathbf{H}_{{i}R_S} \in \mathbb{C}^{{N_{i}}\times {R_S}}$. Furthermore, inter-network interference channels going from primary network to secondary network for $j \in \{1,2\}$ are denoted by $\mathbf{H}_{{R_S}{P_j}} \in \mathbb{C}^{{N_{R_S}}\times {N_{P_j}}}$, and $\mathbf{H}_{{i}{R_P}} \in \mathbb{C}^{{N_{i}}\times {N_{R_P}}}$ for $i \in \{A,B\}$. Moreover, we can define every channel link with the corresponding distance and path-loss exponent ($\tau \geq 2$), i.e., $r_{\!_{m{R_S}}}$ stands for the distance between nodes $m$ and $R_S$, for $m \in \{A,B,P_1,P_2\}$, and the distance between $R_P$ and $i, i \in \{A,B\}$ is described by $r_{\!_{i{R_P}}}$. Accordingly, the path-loss effect on each mentioned channel is defined with $r_{m{R_S}}^{-\tau}$ , and $r_{i{R_P}}^{-\tau}$. Additionally, for $k \in \{A,B,R_S,P_1,P_2,R_P\}$ the transmitted power by node $k$ is presented by $p_k$, $x_k$ is the vector of transmitted i.i.d. symbols from node $k$, and  $d_k$ is the number of data streams demanded by node $k$. Furthermore, $\mathbf{U}_k$ and $\mathbf{V}_k$ present the IA decoding matrix and precoding matrix at node $k$, respectively.

For eliminating the received interference signals at each node, we apply IA. As a result, specific conditions should be satisfied in the network \cite{Zhao2016}. The IA conditions at the secondary relay's receiver in the first and second time-slots are given by \eqref{eq:1} and \eqref{eq:2}.
\begin{align}
  \mathbf{U}^{[j]\emph{H}}_{R_S} \mathbf{H}_{{R_S}{P_j}} \mathbf{V}_{P_j} & =0  \qquad 	\forall j \in \{1,2\}  \label{eq:1}\\
   rank(\mathbf{U}^{[j]\emph{H}}_{R_S} \mathbf{H}_{{R_S}{i}} \mathbf{V}_{i}) & =d_{\!_{i}}, \qquad 	\forall i \in \{A,B\}  \label{eq:2}
\end{align}
in which the superscript $[j]$ indicates time-slot $j$ for $j \in \{1,2\}$. Additionally, the IA conditions at each SU's receiver in the third time-slot are
\begin{align}
  \mathbf{U}^{\emph{H}}_{i} \mathbf{H}_{{i}{R_P}} \mathbf{V}_{R_P} & =0  \qquad 	\forall i \in \{A,B\}   \label{eq:3}\\
   rank(\mathbf{U}^{\emph{H}}_{i} \mathbf{H}_{{i}{R_S}} \mathbf{V}_{R_S}) & =d_{\!_{i}}.  \label{eq:4}
\end{align}
Conditions \eqref{eq:1} and \eqref{eq:3} are responsible for eliminating the interference signals and conditions \eqref{eq:2} and \eqref{eq:4} assure that the desired signal at each node will be recoverable. Moreover, based on \eqref{eq:4}, we can observe that secondary relay designs its precoding matrix $(\mathbf{V}_{R_S})$ in such a way that is suitable for transmitting both SU's signals at time-slot 3.
The secondary relay's received signal at each time slot 1, and 2 is
\begin{equation}\label{eq:5}
\mathbf{y}^{[j]}_{\!_{R_S}}=\sqrt{p_{\!_{i}}r_{\!_{{R_S}{i}}}^{-\tau}}\mathbf{H}_{{R_S}{i}}\mathbf{V}_{i}x_{\!_{i}}+I^{[j]}+\mathbf{n}^{[j]}_{\!_{R_S}},
\end{equation}
in which $i=A$ for $j=1$, and $i=B$ for $j=2$. $\mathbf{n}^{[j]}_{R_S}\sim \mathbb{CN}(0,\mathbf{I}_{N_{R_S}})$ is the additive white gaussian noise (AWGN) at secondary relay's receiver at time-slot $j$. Moreover, $I^{[j]}$ is the received interference at $R_S$ at time-slot $j, j\in \{1,2\}$.
\begin{equation}\label{eq:6}
I^{[j]}= \sqrt{p_{\!_{P_j}}r_{\!_{{R_S}{P_j}}}^{-\tau}}\mathbf{H}_{{R_S}{P_j}}\mathbf{V}_{P_j}\mathbf{x}_{\!_{P_j}}.
\end{equation}

By applying IA decoding matrices at $R_S$ ($\mathbf{U}^{[j]}_{R_S}$, $j \in \{1,2\}$) on its received signals at each time-slot 1 and 2, the interference signals at the relay's receiver will be eliminated. Consequently, the signal to noise ratio (SNR) for secondary relay's received signal from $i$, $i \in \{A,B\}$ can be written by
\begin{equation}\label{eq:7}
  \gamma^{[j]}_{\!_{R_S}}={p_{\!_{i}}}{r^{-\tau}_{\!_{{R_S}{{i}}}}}{{\|\mathbf{U}^{[j]\emph{H}}_{R_S}\mathbf{H}_{R_S{i}}}\mathbf{V}_{i}\|}^2,
\end{equation}
in which $i=A$ for $j=1$, and $i=B$ for $j=2$. The secondary relay decodes and reencodes the SUs' signals $(\tilde{x}_{\!_{A}}, \tilde{x}_{\!_{B}})$, then forwards them in the third time-slot. The forwarded signal by the relay is $\sqrt{p_{\!_{R_S}}}\mathbf{V}_{R_S}\mathbf{x}_{\!_{R_S}}$, in which $\mathbf{x}_{\!_{R_S}}=X_{A}\mathbf{\tilde{x}}_{\!_{A}}+X_{B}\mathbf{\tilde{x}}_{\!_{B}}$, where
\begin{equation}\label{eq:8}
  X_{A}=\frac{1-\theta}{\sqrt{\theta^2+(1-\theta)^2}}, \quad X_{B}=\frac{\theta}{\sqrt{\theta^2+(1-\theta)^2}}.
\end{equation}
$\theta$ is the power control factor which will be used by the secondary relay to allocates its power to each SU's forwarded signal.
The received signal at each SU is given by
\begin{equation}\label{eq:9}
\begin{split}
\mathbf{y}_{\!_{i}}&=\sqrt{p_{\!_{R_S}}r_{\!_{{R_S}{i}}}^{-\tau}}\mathbf{H}_{{i}R_S}\mathbf{V}_{R_S}\mathbf{x}_{\!_{R_S}}\\
&+\sqrt{p_{\!_{R_P}}r_{\!_{i}R_P}^{-\tau}}\mathbf{H}_{{i}R_P}\mathbf{V}_{R_P}\mathbf{x}_{\!_{R_P}}+\mathbf{n}_{\!_{i}},\\
&\forall i \in \{A,B\}.
 \end{split}
\end{equation}
where $\mathbf{n}_{\!_{i}} \sim \mathbb{CN}(0,\mathbf{I}_{N_{S_i}})$ is AWGN at SUs receivers. SUs ($A$ and $B$) provide their energy by WEH using the PS protocol. It means that their received signal will be divided into two parts by their PS ratio $(\rho_{\!_{i}}, i \in \{A,B\})$ . One part will be used for power harvesting and providing their necessary energy. The other part goes to the IP unit for decoding their desired information (as can be seen in Figure \ref{fig:1}). The harvested power at each SU in the $3^{rd}$ time-slot is given by
\begin{equation}\label{eq:10}
\begin{split}
   p_{\!_{i}}&=\eta (1-\rho_{\!_{i}})[p_{\!_{R_S}} r_{\!_{{R_S}{i}}}^{-\tau} {\|\mathbf{H}_{{i}R_S}\mathbf{V}_{R_S}\|}^2 \\
     &+ p_{\!_{R_P}} r_{\!_{{R_P}{i}}}^{-\tau} {\|\mathbf{H}_{{i}R_P}\mathbf{V}_{R_P}\|}^2],\forall i \in \{A,B\},
\end{split}
\end{equation}
in which $\eta$ is the power conversion efficiency. It is worth mentioning that since the amount of harvested power from AWGN is not considerable, can be ignored in \eqref{eq:10}. At its IP unit, since each SU is aware of its own transmitted signal in the past time-slots, it subtracts this signal from all received signals. Then, by applying the IA decoding matrix eliminates the interference signals coming from the primary network. As a result, the received signal at the IP unit of $i$ for $i \in \{A,B\}$ can be written by
\begin{equation}\label{eq:11}
\begin{split}
&\mathbf{y}^{\!_{IP}}_{\!_{i}}=\sqrt{p_{\!_{R_S}}{r_{\!_{{R_S}{i}}}^{-\tau}}{\rho_{\!_{i}}}}X_{{\hat{i}}}\mathbf{U}^{\emph{H}}_{i} \mathbf{H}_{{i}{R_S}}\mathbf{V}_{R_S} \mathbf{x}_{\!_{{\hat{i}}}}+ {\rho_{\!_{i}}}\mathbf{n}_{\!_{i}}\\
&\forall i \in \{A,B\},
\end{split}
\end{equation}
where {\color{black}if} $i=A$, $\hat{i}=B$ and if $i=B$, $\hat{i}=A$. Therefore, the SNR at each SU's receiver is
\begin{equation}\label{eq:12}
 \gamma_{\!_{i}}={p_{\!_{R_S}}}{r^{-\tau}_{\!_{{R_S}{i}}}}{\rho_{\!_{i}}}{X^{2}_{{\hat{i}}}}{\|\mathbf{U}^{\emph{H}}_{i}\mathbf{H}_{{i}R_S}\mathbf{V}_{R_S}\|}^2.
\end{equation}

\section{OUTAGE PROBABILITY MINIMIZATION}\label{section.solution}
The system outage for the secondary network in proposed scheme consisting of a DF relay happens when the data rate of each one of the four links (from SUs to relay and relay to SUs) falls bellow a rate threshold \cite{Shi2019}. Consequently, for a given SNR threshold $\gamma_{\!_{th}}$ , the secondary network overall outage probability $(\mathbb{P}_{out})$ is
\begin{equation}\label{eq:13}
\begin{split}
 &\mathbb{P}_{out}=1-\mathbb{P}_{sucess}\\
 &=1-\mathbb{P}(\gamma^{[1]}_{\!_{R_S}}\geq \gamma_{\!_{th}},\gamma^{[2]}_{\!_{R_S}}\geq \gamma_{\!_{th}},\gamma_{\!_{A}}\geq \gamma_{\!_{th}},\gamma_{\!_{B}}\geq \gamma_{\!_{th}})
 \end{split}
\end{equation}
where $\mathbb{P}(.)$ denotes the probability, and $\mathbb{P}_{sucess}$ is the probability of successful transmission in all four links. If the secondary relay fails to decode the transmitted signal of SUs successfully, the occurrence of the outage will be definite. Hence, we assume that $\gamma^{[1]}_{\!_{R_S}}\geq \gamma_{\!_{th}}$ and $\gamma^{[2]}_{\!_{R_S}}\geq \gamma_{\!_{th}}$ are true to make sure that the relay decodes the SU signals correctly. Now, we should focus on satisfying $\gamma_{\!_{A}}\geq \gamma_{\!_{th}}$  and $\gamma_{\!_{B}}\geq \gamma_{\!_{th}}$. For this goal, we attempt to maximize the lowest value between $\gamma_{\!_{A}}$ and $\gamma_{\!_{B}}$ (finding $max\; min (\gamma_{\!_{A}},\gamma_{\!_{B}})$ . Thus, we have the following optimization problem.
\begin{subequations}
\begin{align}
   \underset {\!_{\mathbf{U}_k,\mathbf{V}_k},k \in \{A,B,R_S\}\rho_{\!_{A}}, \rho_{\!_{B}}, \theta} {\textrm{max}}  \; &\textrm{min}(\gamma_{\!_{A}},\gamma_{\!_{B}})  \label{eq:14.a} \\
  \textrm{s.t. }\qquad &\gamma^{[1]}_{\!_{R_S}}\geq \gamma_{\!_{th}} \label{eq:14.b} \\
  &\gamma^{[2]}_{\!_{R_S}}\geq \gamma_{\!_{th}} \label{eq:14.c}\\
  &0\leq \rho_{\!_{i}} \leq 1, i\in \{A,B\}  \label{eq:14.d}\\
  &0<\theta<1 \label{eq:14.e} \\
&\eqref{eq:1}-\eqref{eq:4}  \label{eq:14.f}
\end{align}
\end{subequations}
To solve the above design problem, we propose our solution in two following steps.

\subsection{Optimizing the PS ratios and power control factor}\label{subsec1}
In this part, we set constant values for IA beamforming matrices and make attempt to choose the optimal PS ratios and power control factor at SUs and relay, respectively. For one, We have the following design problem to achieve $\rho_{\!_{i}}$, $i \in \{A,B\}$.
\begin{subequations}
  \begin{align}
   \underset {\!_{\rho_{\!_{A}}, \rho_{\!_{B}}}} {\textrm{max}}  \; &\textrm{min}(\gamma_{\!_{A}},\gamma_{\!_{B}})  \label{eq:15.a} \\
 \textrm{ s.t.} \quad &\gamma^{[1]}_{\!_{R_S}}\geq \gamma_{\!_{th}} \label{eq:15.b} \\
  &\gamma^{[2]}_{\!_{R_S}}\geq \gamma_{\!_{th}} \label{eq:15.c}\\
  &0\leq \rho_{\!_{i}} \leq 1   \label{eq:15.d}
  \end{align}
\end{subequations}
According to \eqref{eq:7}, \eqref{eq:10}, \eqref{eq:15.b} and \eqref{eq:15.c}, with a little calculation, we have $\rho_{\!_{i}}\leq 1-\frac{\gamma_{\!_{th}}}{Z_{i}}$, $i \in \{A,B\}$, where
\begin{equation}\label{eq:16}
\begin{split}
 Z_{i}&={r^{-\tau}_{\!_{{R_S}{{i}}}}}\eta{\|\mathbf{U}^{[j]\emph{H}}_{R_S}\mathbf{H}_{{R_S}{i}}\mathbf{V}_{i}\|^2}\\
 &\times [p_{\!_{R_S}} r_{\!_{{R_S}{i}}}^{-\tau} {\|\mathbf{H}_{{i}R_S}\mathbf{V}_{R_S}\|}^2 \\
     &+ p_{\!_{R_P}} r_{\!_{{R_P}{i}}}^{-\tau} {\|\mathbf{H}_{{i}R_P}\mathbf{V}_{R_P}\|}^2].
     \end{split}
\end{equation}
 Now, we can rewrite (15) as follows.
\begin{subequations}
  \begin{align}
   \underset {\!_{\rho_{\!_{A}}, \rho_{\!_{B}}}} {\textrm{max}}  \; &\textrm{min}(\gamma_{\!_{A}},\gamma_{\!_{B}})  \label{eq:17.a} \\
 \textrm{ s.t.} \quad &0\leq \rho_{\!_{i}}\leq 1-\frac{\gamma_{\!_{th}}}{Z_{i}} \label{eq:17.b}.
  \end{align}
\end{subequations}
Based on \eqref{eq:12}, with increasing in $\rho_{\!_{A}}$ and $\rho_{\!_{B}}$ , $\gamma_{\!_{A}}$  and $\gamma_{\!_{B}}$  will increase, respectively. Therefore, the optimal values for the PS ratios at each SU ($\rho_{\!_{A}}^{*}$ and $\rho_{\!_{B}}^{*}$) is the maximum value possible within their intervals. It means that
\begin{equation}\label{eq:18}
  \rho_{\!_{i}}^{*}=\textrm{max}(1-\frac{\gamma_{\!_{th}}}{Z_{i}},0).
\end{equation}
Considering \eqref{eq:18}, we can make the following observations. Firstly, the proposed closed-form solution for optimal PS ratios at SUs is dynamic, which means that PS ratios will be chosen according to the instantaneous channel state information (CSI). This feature, makes the proposed scheme more flexible and practical. Secondly, choosing $\rho_{\!_{i}}^{*}$ is independent of relay’s power control factor ($\theta$), so aiming to find $\theta$, we can substitute the calculated optimal PS ratios of SUs in the objective function of (14) to reach
\begin{subequations}
\begin{align}
   \underset {\theta} {\textrm{max}}  \quad &\textrm{min}(\frac{{Y_B}{\theta^2}}{\theta^2+(1-\theta)^2},\frac{{Y_A}{(1-\theta)^2}}{\theta^2+(1-\theta)^2})  \label{eq:19.a} \\
  \textrm{ s.t.} \quad  &0<\theta<1 \label{eq:19.b}
\end{align}
\end{subequations}
where, $Y_i=p_{\!_{R_S}}r_{\!_{{R_S}{i}}}^{-\tau}\|\mathbf{U}^{\emph{H}}_{i}\mathbf{H}_{{i}R_S}\mathbf{V_{R_S}}\|^2$, $i \in \{A,B\}$. Now we have two following scenarios for solving (19).\\
\textbf{Scenario 1:} \, $\frac{{Y_A}{(1-\theta)^2}}{\theta^2+(1-\theta)^2}\geq \frac{{Y_B}{\theta^2}}{\theta^2+(1-\theta)^2}$, it means that $r_{\!_{{R_S}{A}}}^{-\tau} \rho_{\!_{A}}^{*} {\|\mathbf{U}^{\emph{H}}_{A}\mathbf{H}_{{A}R_S}\mathbf{V}_{R_S}\|}^2 (1-\theta)^2 \geq r_{\!_{{R_S}{B}}}^{-\tau} \rho_{\!_{B}}^{*} {\|\mathbf{U}^{\emph{H}}_{B}\mathbf{H}_{{B}R_S}\mathbf{V}_{R_S}\|}^2 \theta^2$, so we can rewrite (19) as
\begin{subequations}
\begin{align}
   \underset {\theta} {\textrm{max}}  \quad &(\frac{{Y_B}{\theta^2}}{\theta^2+(1-\theta)^2})  \label{eq:20.a} \\
  \textrm{ s.t.} \quad  &0<\theta \leq Q \label{eq:20.b}
\end{align}
\end{subequations}
in which
\begin{equation}\label{eq:21}
\textstyle Q=\frac{\sqrt{{r_{\!_{{R_S}{A}}}^{-\tau}}{\rho_{\!_{A}}}}\mathbf{U}^{\emph{H}}_{A} \mathbf{H}_{{A}{R_S}}\mathbf{V}_{R_S}}{\sqrt{{r_{\!_{{R_S}{A}}}^{-\tau}}{\rho_{\!_{A}}}}\mathbf{U}^{\emph{H}}_{A} \mathbf{H}_{{A}{R_S}}\mathbf{V}_{R_S}+\sqrt{{r_{\!_{{R_S}{B}}}^{-\tau}}{\rho_{\!_{B}}}}\mathbf{U}^{\emph{H}}_{B} \mathbf{H}_{{B}{R_S}}V_{R_S}}.
\end{equation}
Since with increasing in $\theta$ the objective function in (20) increases, the optimal solution of (20) is $\theta^*=Q$.\\
\textbf{Scenario 2:}\, $\frac{{Y_A}{(1-\theta)^2}}{\theta^2+(1-\theta)^2}\leq \frac{{Y_B}{\theta^2}}{\theta^2+(1-\theta)^2}$, which means $r_{\!_{{R_S}{A}}}^{-\tau} \rho_{\!_{A}}^{*} {\|\mathbf{U}^{\emph{H}}_{A}\mathbf{H}_{{A}R_S}\mathbf{V}_{R_S}\|}^2 (1-\theta)^2 \leq r_{\!_{{R_S}B}}^{-\tau} \rho_{\!_{B}}^{*} {\|\mathbf{U}^{\emph{H}}_{B}\mathbf{H}_{{B}R_S}\mathbf{V}_{R_S}\|}^2 \theta^2$. Consequently, for this case, we can rewrite (19) as follows.
\begin{subequations}
\begin{align}
   \underset {\theta} {\textrm{max}}  \quad &(\frac{{Y_A}{(1-\theta)^2}}{\theta^2+(1-\theta)^2})  \label{eq:22.a} \\
\textrm{ s.t.} \quad    &Q\leq \theta < 1 \label{eq:22.b}
\end{align}
\end{subequations}
We can observe that with increasing of $\theta$, \eqref{eq:22.a} decreases. Therefore the optimal solution for $\theta$  in this scenario again is $\theta^*=Q$.
\subsection{Managing the interference signals with IA}\label{subsec2}
We design the IA beamforming matrices by adapting the MMSE iterative algorithm applied in \cite{Aquilina2017} and \cite{soltani2021performance} to make a trade-off between eliminating the interferences and minimizing the possibility of happening an error. Therefore, we have the following design problem for choosing IA decoding matrix for the secondary relay at the first and second time-slots.
\begin{equation}\label{eq:23}
\underset {\mathbf{U}_{R_S}^{[j]}}{min} \quad E\{\|\mathbf{f}_{\!_{R_S}}^{[j]}-\mathbf{x}_{\!_{i}}\|^2\}
\end{equation}
in which $i=A$ for $j=1$, and $i=B$ for $j=2$, and
\begin{equation}\label{eq:24}
\begin{split}
  \mathbf{f}_{\!_{R_S}}^{[j]}&=\sqrt{p_{\!_{i}}r_{\!_{{R_S}{i}}}^{-\tau}}\mathbf{U}_{R_S}^{[j]\emph{H}}\mathbf{H}_{{R_S}{i}}\mathbf{V}_{i}\mathbf{x}_{{i}}\\
  &+\sqrt{p_{\!_{P_j}}r_{\!_{{R_S}{P_j}}}^{-\tau}}\mathbf{U}_{R_S}^{[j]\emph{H}}{\mathbf{H}_{{R_S}{P_j}}}\mathbf{V}_{P_j}\mathbf{x}_{\!_{P_j}}\\
  &+\mathbf{U}_{R_S}^{[j]\emph{H}}\mathbf{n}^{[j]}_{\!_{R_S}}.
\end{split}
\end{equation}
 By solving \eqref{eq:23}, we can achieve the secondary relay's decoding matrix at each iteration, which is written by %\eqref{eq:25} on the top of the next page.
  \begin{figure*}[h]
\begin{align} \label{eq:25}
  &\mathbf{U}_{R_S}^{[j]}= \\ \nonumber
  &\frac{\sqrt{p_{{i}}r_{\!_{{R_S}{i}}}^{-\tau}}\mathbf{H}_{{R_S}{i}}\mathbf{V}_{i}}{{p_{{i}}r_{\!_{{R_S}{i}}}^{-\tau}}\mathbf{H}_{{R_S}{i}}\mathbf{V}_{i}\mathbf{H}_{{R_S}{i}}^{\emph{H}}\mathbf{V}_{i}^{\emph{H}}
  +{p_{\!_{P_j}}r_{\!_{{R_S}{P_j}}}^{-\tau}}\mathbf{H}_{{R_S}{P_j}}\mathbf{V}_{P_j}\mathbf{H}_{{R_S}{P_j}}^{\emph{H}}\mathbf{V}_{P_j}^{\emph{H}}+\mathbf{I}_{N_{R_S}}} 
\end{align}
\end{figure*}\\
The detailed calculation related to \eqref{eq:25} can be found in Appendix. The same procedure as for \eqref{eq:25} can be proceeded for achieving the decoding matrices of primary relay at the first and second time-slots. The MMSE IA algorithm works based on channels' reciprocity. Therefore, after choosing the proper decoding matrices, we will consider the reciprocal network for the first two time-slots to find the appropriate IA precoding matrices of SUs and PUs. In the reciprocal network we have $\overleftarrow{\mathbf{H}_{ab}}=\mathbf{H}_{ba}^{\emph{H}}$ , so we assume the transmission direction is reversed, and the calculated decoding matrices act as precoding matrices. We skip the detailed calculation of the precoding matrices at the first two time-slots (SUs and PUs' precoding matrices) since the process is the same as choosing \eqref{eq:25}. Here, we consider the third time-slot, when relays forward the information signals. At the third time-slot, decoding matrices at SUs ($\mathbf{U}_{i}$ for $i \in \{A,B\}$) and PUs ($\mathbf{U}_{P_j}$ for $j \in \{1,2\} $) can be found again in the similar way as for \eqref{eq:25}. Now, for designing the precoding matrix of the secondary relay at the third time-slot, in the reciprocal network we have the following problem ($\overleftarrow{A}$ indicates that the parameter $A$ is in the reciprocal network).
\begin{equation}\label{eq:26}
\underset {\mathbf{V_{R_S}}}{min} \quad E\{\|\mathbf{\overleftarrow{f}}_{\!_{R_S}}-(X_{B}\mathbf{x}_{\!_{A}}+X_{A}\mathbf{x}_{\!_{B}})\|^2\}
\end{equation}
where
\begin{equation}\label{eq:27}
\begin{split}
 \mathbf{\overleftarrow{f}}_{\!_{R_S}}&={\sum_{i=A,B}}\sqrt{p_{{i}}r_{\!_{{R_S}{i}}}^{-\tau}}X_{{\hat{i}}}\mathbf{V}_{R_S}^{\emph{H}}\overleftarrow{\mathbf{H}}_{{{R_S}{S_i}}}\mathbf{U}_{i}\mathbf{x}_{\!_{i}}\\
  &+\overset{2}{\sum_{j=1}}\sqrt{p_{\!_{P_j}}r_{\!_{{R_S}{P_j}}}^{-\tau}}\mathbf{V}_{R_S}^{\emph{H}}\overleftarrow{\mathbf{H}}_{{{R_S}{P_j}}}\mathbf{U}_{P_j}\mathbf{x}_{\!_{P_j}}\\
  &+\mathbf{V}_{R_S}^{\emph{H}}\mathbf{\overleftarrow{\mathbf{n}}}_{\!_{R_S}}.
 \end{split}
\end{equation}
Finally, with the same calculations as in the Appendix, the secondary relay's precoding matrix for intended signals of SUs at the third time-slot will be achieved as %The result is written on the top of the next page.

 \begin{figure*}[h]
\begin{align}  \label{eq:28}
 &\mathbf{V}_{R_S}= \\ \nonumber
 & { \small \frac{{{\sum_{i=A,B}}}\sqrt{p_{{i}}r_{\!_{{R_S}{i}}}^{-\tau}}X_{{\hat{i}}}\overleftarrow{\mathbf{H}}_{{{R_S}{i}}}\mathbf{U}_{i}}
  {{\sum_{i=A,B}}p_{{i}}r_{\!_{{R_S}{i}}}^{-\tau}{X_{{\hat{i}}}}^2\overleftarrow{\mathbf{H}}_{{{R_S}{i}}}\mathbf{U}_{i}{\mathbf{U}^{\emph{H}}_{i}}\overleftarrow{\mathbf{H}}^{\emph{H}}_{{{R_S}{i}}}
 +\overset{2}{\sum_{j=1}}p_{\!_{P_j}}r_{\!_{{R_S}{P_j}}}^{-\tau}\overleftarrow{\mathbf{H}}_{{{R_S}{P_j}}}\mathbf{U}_{P_j}\mathbf{U}^{\emph{H}}_{P_j}\overleftarrow{\mathbf{H}}^{\emph{H}}_{{{R_S}{P_j}}} +\mathbf{I}_{N_{R_S}}}}
\end{align}
\vspace{8pt}
%\hrule
\end{figure*}

An abridge to the proposed two-step solving method presented in Algorithm 1.\\

%% Algorithm 
\begin{algorithm}
\caption{  The proposed two-step outage minimization algorithm}
\label{alg:alg}
\begin{algorithmic}[1]
\State \textit{Input:} all channel matrices $\mathbf{H}$, path-loss effects $r^{-\tau}$, powers $p$, $\gamma_{\!_{th}}$, $maxiter$, $maxiter\emph{2}$ 
\State \textit{Initialization:} all IA matrices $(\mathbf{U,V})$ are randomly initialized.
\While{$iter < maxiter$}
\State Calculate $\rho^{*}_{{i}}$ using \eqref{eq:18}.
\State Choose $\theta^{*}$ using \eqref{eq:21}.
\While{$iter\emph{2} < maxiter\emph{2}$}
\State Choose $\mathbf{U}^{[j]}_{R_S},\mathbf{U}^{[j]}_{R_P}$, for $j \in \{1,2\}$, using \eqref{eq:25} and Appendix.
\State Update $\mathbf{V_{i},V_{P_j}}$, for $i \in \{A,B\}, j \in \{1,2\}$ similar to \eqref{eq:28}.
\State Calculate $\mathbf{U}_{i},\mathbf{U}_{P_j}$, for $i \in \{A,B\}, j \in \{1,2\}$ similar to \eqref{eq:25}.
\State Update $\mathbf{V}_{R_S},\mathbf{V}_{R_P}$ using \eqref{eq:28}.
\State $iter\emph{2} = iter\emph{2}+1$
\EndWhile
\State $iter = iter+1$
\EndWhile
\end{algorithmic}
\end{algorithm}

\section{NUMERICAL RESULTS}\label{section.results}
In this section, the performance of our proposed scheme for the communication between SUs over TW relay with optimal PS ratios and power control selection together with applying IA at all nodes for eliminating interference signals will be evaluated and compared to the benchmark schemes, namely, the static equal PS schemes, the scheme with maximum ratio transmission (MRT)-maximum ratio combining (MRC) beamformers \cite{hoang2023secrecy}, and the static power control scheme. In the static equal PS schemes, we apply IA matrices with the proposed MMSE approach at all nodes, and for wireless-powered SUs, we consider that $\rho_{\!_{A}}=\rho_{\!_{B}}$ is set as $0.3$, $0.5$ and $0.7$; also, $\theta$ is calculated with the proposed optimal solution in \eqref{eq:21}. Furthermore, for the scheme with MRT-MRC beamformers, we apply MRT matrices at transmitters and MRC matrices at receivers, instead of IA beamformers. It is worth mentioning that MRT-MRC beamforming is optimal for MIMO systems in the absence of interference signals \cite{nguyen2020analysis}. Also, in this scheme, we consider optimal values for $\rho_{\!_{A}}$, $\rho_{\!_{B}}$ and $\theta$ according to our proposed scheme. Moreover, In the static power control scheme, it is assumed that the secondary relay allocates half of its transmit power to each SU $(\theta=0.5)$, and the PS ratios of SUs are optimally chosen based on \eqref{eq:18}. Also, in this scheme, IA matrices are applied at all nodes according to the proposed MMSE algorithm. We consider $p_{\!_{P_j}}=p_{\!_{R_P}}=35dBm$, $j \in \{1,2\}$, $\tau=2.7$, $\eta=0.8$, and $d_{\!_k}=1$, for $k \in \{A,B,R_S,R_P,P_1,P_2\}$ (single data stream). In addition, we consider $r_{\!_{P_1P_2}}=r_{\!_{AB}}=1meter$, $r_{\!_{P_1R_P}}=r_{\!_{R_PP_2}}=0.5meter$ (normalized distances), and $r_{\!_{R_SR_P}}=2meter$. Also, unless otherwise stated, we set $N_k=4$ for $k \in \{A,B,R_S,P_1,P_2,R_P\}$, $p_{\!_{R_S}}=20dBm$, $\gamma_{\!_{th}}=1dB$, and $r_{\!_{AR_S}}=r_{\!_{R_SB}}=0.5meter$. It is worth mentioning that the remaining distances will be calculated using Pythagorean theorem (for instance, the distance between $R_P$ and $A$ is $\sqrt{{r^{2}_{\!_{R_SR_P}}}+r_{\!_{AR_S}}^{2}}$). {Table \ref{Table1} summarizes the default values of the parameters that are used in the simulations.}

% Table 1 %%%%%%%%%%%%%%%%%%%%%%%%%%%%%%%%%%%%%%%%%%%%%%%%%%%%%%%%%%%%%%%%%

\begin{table}

\centering
\caption{Default simulation parameter settings}
\begin{tabular}{ |p{1.9cm}|p{1.8cm}||p{1.9cm}|p{1.8cm}||p{1.9cm}|p{1.8cm}|}
 \hline
 Parameter&Value&Parameter&Value&Parameter&Value\\
 \hline
  $\tau$ & 2.7 &   $\eta$ &  0.8 & $d_{\!_k}$&   1\\
  
  $\gamma_{\!_{th}}$ & $1\ dB$  &   $N_k$ &  4 & $p_{\!_{R_S}}$&   $20 \ dBm$ \\
  
  $p_{\!_{R_P}}$ & $35 \ dBm$ &   $p_{\!_{P_1}}$ & $35 \ dBm$  &$p_{\!_{P_2}}$ & $35 \ dBm$ \\
  
  $r_{\!_{AR_S}}$ & $0.5 \ meter$ &   $r_{\!_{R_SB}}$ & $0.5 \ meter$ & $r_{\!_{R_SR_P}}$ & $2 \ meter$\\
  
  $r_{\!_{P_1R_P}}$ & $0.5 \ meter$ &   $r_{\!_{R_PP_2}}$ & $0.5 \ meter$ & $r_{\!_{R_SP_1}}$ & $2.06 \ meter$\\

  $r_{\!_{R_SP_2}}$ & $2.06 \ meter$ &   $r_{\!_{R_PA}}$ & $2.06 \ meter$ & $r_{\!_{R_PB}}$ & $2.06 \ meter$\\
 \hline
\end{tabular}
\centering
\label{Table1}
\end{table}

%%%%%%%%%%%%%%%%%%%%%%%%%%%%%%%%%%%%%%%%%%%%%%%%%%%%%%%%%%%%%%%%%%%

\begin{figure}[t]
  \centering
  \includegraphics[scale=0.55, angle=0]{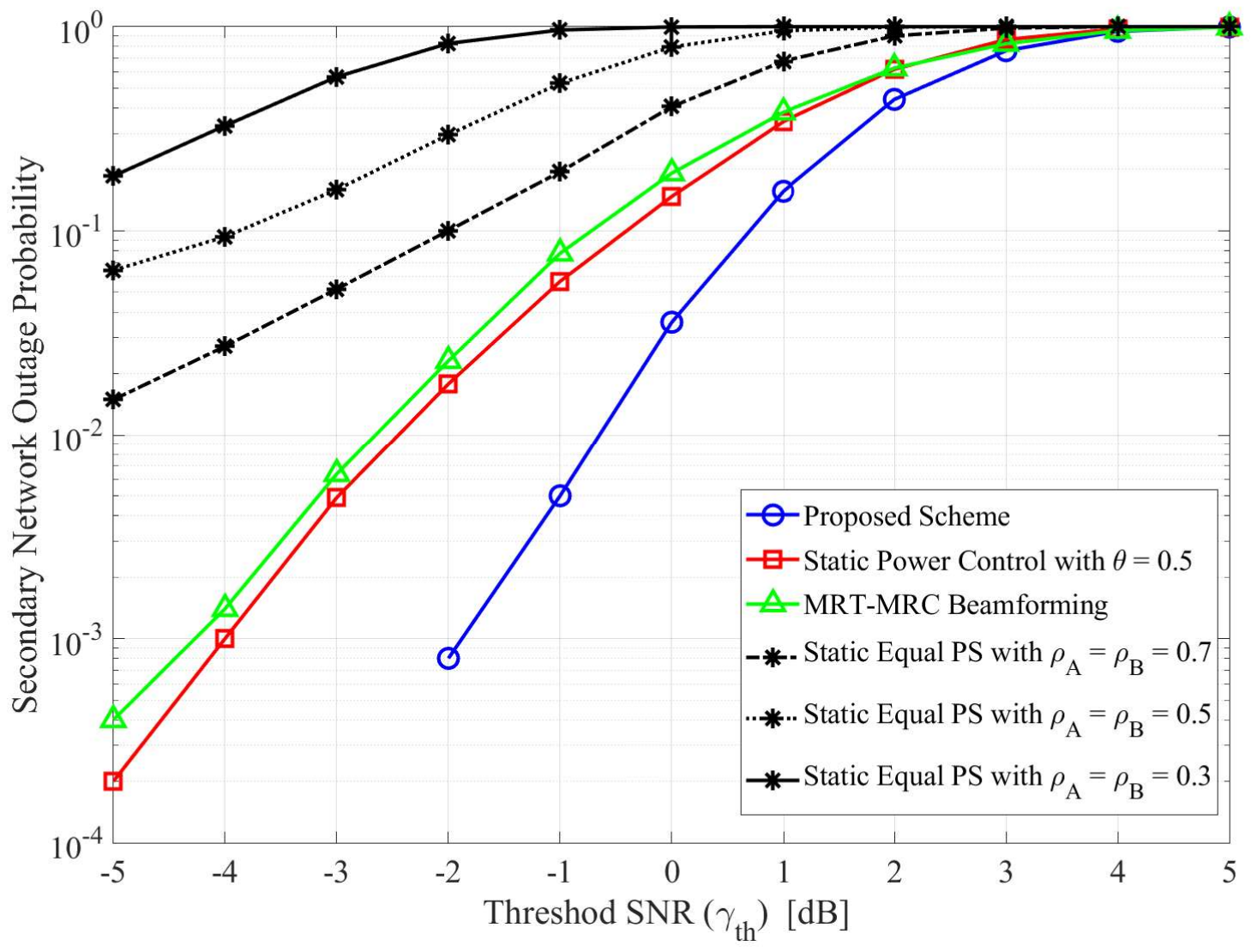}
  \caption{Secondary network outage probability versus the threshold SNR $(\gamma_{\!_{th}})$ }\label{fig:2}
\end{figure}

One of the key factors for evaluating the outage performance is the threshold SNR $(\gamma_{\!_{th}})$. Figure \ref{fig:2} presents the outage performance of the proposed scheme and benchmark schemes versus  $\gamma_{\!_{th}}$. It goes without saying that the outage probability increases with elevating $\gamma_{\!_{th}}$. We can witness that the proposed scheme outperforms the benchmark schemes since it chooses optimal PS ratios and power control factor based on the instantaneous CSI together with removing the negative effect of interferences. {For instance, when the threshold SNR is set to $\gamma_{\!_{th}}=-2dB$, the proposed scheme achieves an improvement of at least one order of magnitude reduction in the outage probability compared to the benchmark methods.} The superiority of the static power control scheme with $\theta=0.5$ over the static equal PS scheme with $\rho_{\!_{A}}=\rho_{\!_{B}}=0.5$ presents that the effect of choosing optimal PS ratios at SUs is higher than optimally dividing power at the relay. Moreover, the proposed scheme’s performance gain by applying IA over the scheme with MRT-MRC filters indicates the importance of removing the interferences in the secondary relay and IP unit of wireless-powered SUs in the system.

\begin{figure}[t]
  \centering
  \includegraphics[scale=0.55, angle=0]{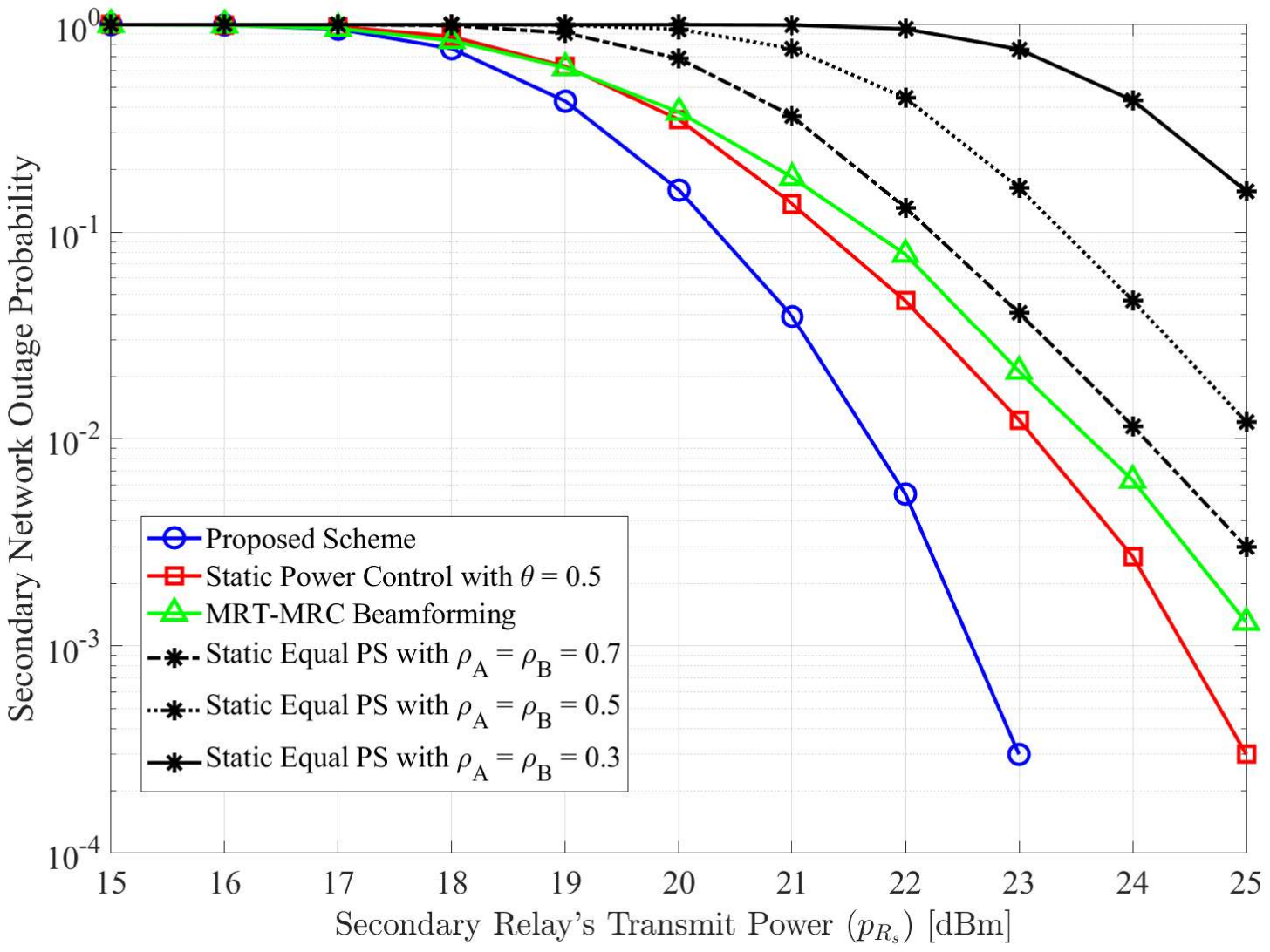}
  \caption{Secondary network outage probability versus secondary relay's transmit power $(p_{\!_{R_S}})$ }\label{fig:3}
\end{figure}

 We illustrate the outage performance of our proposed scheme comparing to the benchmark schemes versus the secondary relay’s transmit power in Figure \ref{fig:3}. As we expected, the outage probability decreases with increasing the relay’s transmit power since it not only does heighten the quality of SUs’ desired signals but also gives more power to them for harvesting. {The proposed scheme requires the secondary relay to use at least $2dBm$ less power than the benchmark schemes in order to achieve an outage probability of $10^{-3}$.} The figure exposes the optimality of our proposed closed-form solution for SUs’ PS ratios over applying the conventional static PS ratios ($\rho_i=0.3$, $\rho_i=0.5$, and $\rho_i=0.7$ for $i \in \{A,B\}$) since it chooses the PS ratio based on the instantaneous CSI aiming for achieving the lowest outage probability. Furthermore, our proposed scheme’s superiority over the static power control scheme indicates that allocating half of the transmit power to each SU by the relay is not the best solution for the secondary network. Consequently, as can be seen, implementing the proposed power control strategy at the secondary relay can make a considerable improvement. Moreover, we can observe that the performance of the scheme with MRT-MRC filters is considerably lower than our proposed scheme, which shows the necessity of managing the interference signals in the network. Hence, one of the claims we can make based on Figure \ref{fig:3} is that exerting a strong interference management technique like IA together with WEH is needed to achieve the best performance in the wireless-powered SUs underlaying CR network.

 \begin{figure}[t]
  \centering
  \includegraphics[scale=0.55, angle=0]{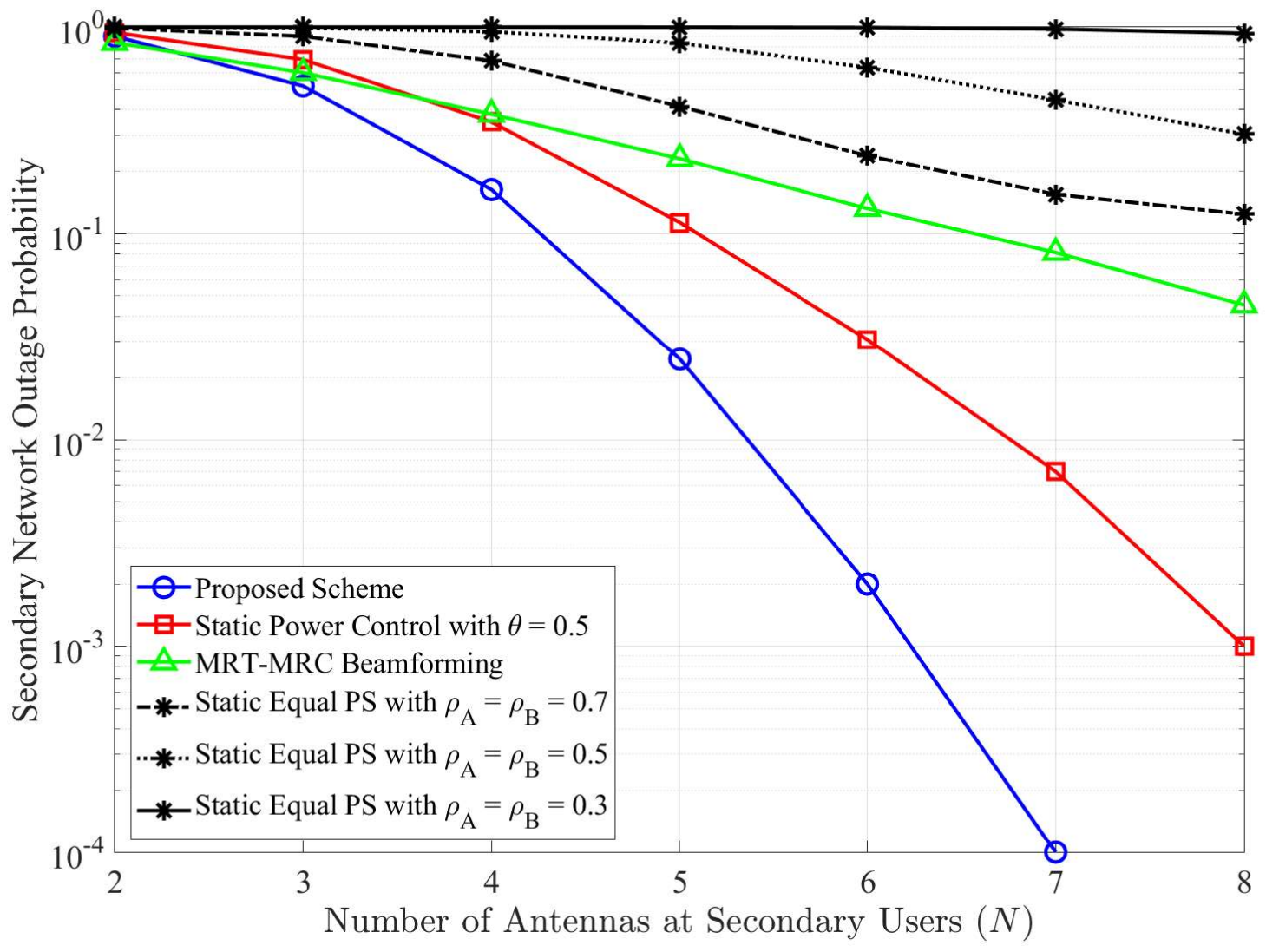}
  \caption{Secondary network outage probability versus the number of antennas at SUs}\label{fig:4}
\end{figure}
Figure \ref{fig:4} presents the outage probability of secondary network versus the number of antennas at SUs $(N_{A}=N_{B}=N)$. As the number of antennas at the wireless-powered SUs increases, the outage probability of the secondary network decreases because of two main reasons: 1. elevating the degrees of freedom, which results in better interference mitigation, and 2. growing the amount of harvested power at each SU. Also, it can be observed that when SUs are equipped with two antennas, the performance of the scheme with MRT-MRC beamformers is slightly better than the proposed scheme. The reason is that with low antenna numbers, SUs cannot optimally manage joint harvesting power and eliminating interferences using IA, as for applying IA successfully, there should be enough degrees of freedom at each node \cite{Zhao2016}. Again in this figure, the overall excellence of the proposed scheme over the benchmark schemes is visible. {As an illustration, in the proposed scheme, the nodes require $N=6$ antennas to achieve an outage probability of approximately $10^{-4}$. However, in the static power control scheme, $N=8$ antennas are needed to achieve the same performance, and the other benchmarks require even more antennas. Therefore, the proposed scheme offers a lower number of antennas, which reduces the cost and complexity of the nodes.}

 \begin{figure}[t]
  \centering
  \includegraphics[scale=0.55, angle=0]{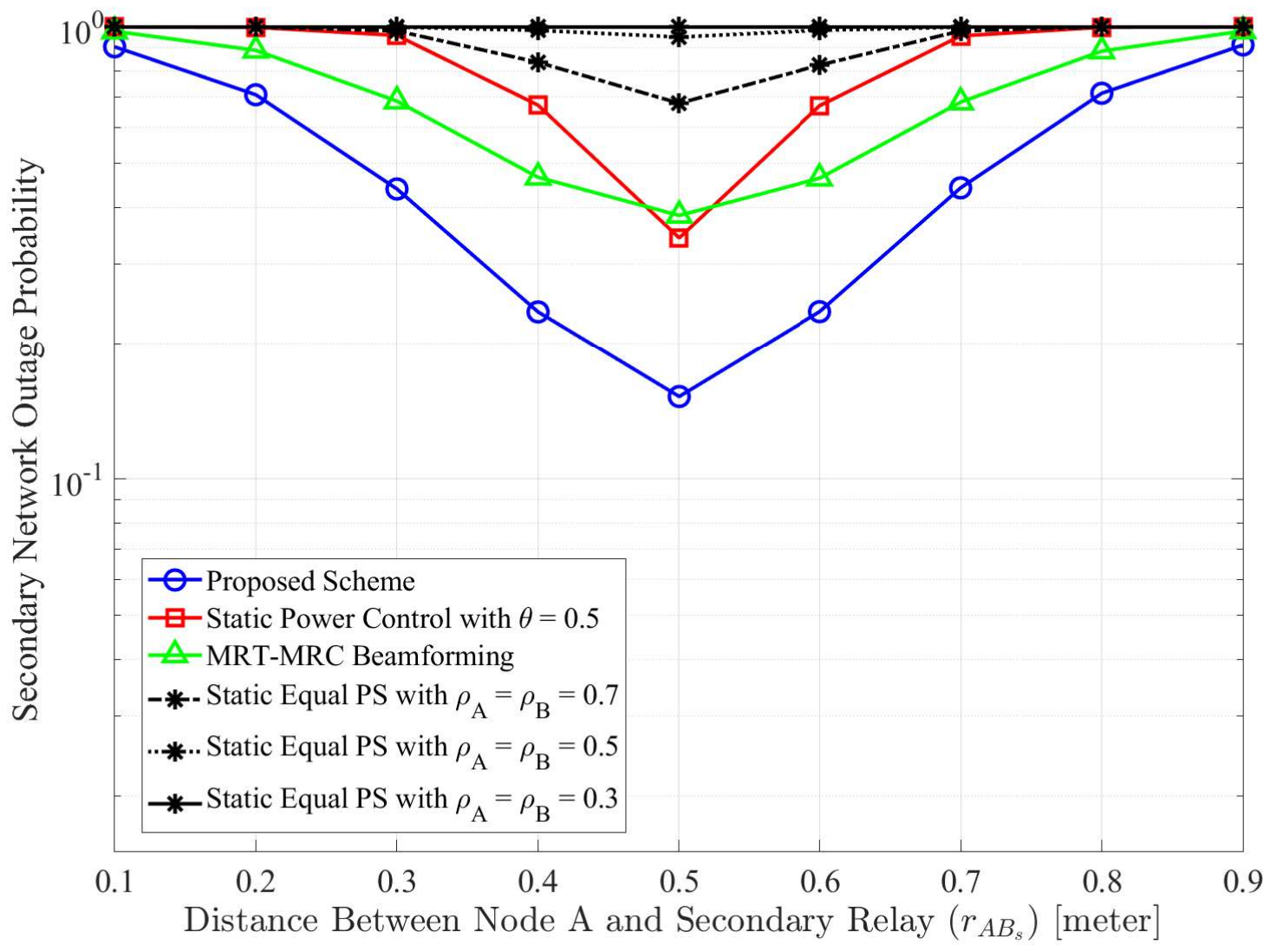}
  \caption{Secondary network outage probability versus the distance between $A$ and $R_S$}\label{fig:5}
\end{figure}

 In Figure \ref{fig:5}, the outage probability of secondary network versus the distance of $R_S$ from SU $A$ $(r_{\!_{{R_S}{A}}})$ is presented. We assume that all nodes except for the secondary relay are at a standstill, and $R_S$ moves between SUs starting at the closest point to $A$ and finishing near $B$. We can observe that the secondary network’s outage probability is minimized when the secondary relay is placed almost in the middle of the distance between SUs. {As a case in point, in the proposed scheme, when the secondary relay is positioned in the middle of the SUs, the outage probability is approximately $10^{-1}$.} As $R_S$ gets closer to one SU, because of intensifying the path-loss effect, the other SU’s outage probability, and consequently, according to \eqref{eq:13}, the outage probability of secondary network grows. The path-loss effect on wireless-powered nodes not only does cause the quality of the desired signal to fall down but also reduces the amount of power for the node to harvest. Therefore, the path-loss effect is more intense on wireless powered nodes comparing to regular nodes. Furthermore, we can witness the importance of controlling power at $R_S$ in this figure by observing the better overall performance gain of the scheme with MRT-MRC beamformers over the scheme with static power control factor. Besides, the performances of schemes with static equal PS ratios at SUs are lowest at any given distance as they cannot optimally manage their dedicated portions of the received signals to EH and IP when the effect of path-loss changes. After all, based on this figure, we can again observe that the proposed scheme outperforms the
benchmark schemes.

%After all, based on this figure, we can observe that the proposed scheme considerably outperforms the benchmark schemes.

 \section{Conclusion}\label{section.conclusion}
 In this paper, we investigated the performance of wireless-powered SUs communicating over a TW DF relay underlaying cognitive radio network in the presence of PUs and their dedicated relay. We formulated our design problem to minimize the SUs outage probability. Then, we proposed our solution in two steps. Firstly, we presented closed-form solutions for PS ratios at each SU and the secondary relay's power control factor. Secondly, we calculated the IA beamforming matrices for avoiding the interference between SUs and PUs in an iterative manner. Through simulations, we demonstrated the advantages of our proposed scheme over benchmark schemes. All in all, in the underlay communication between SUs in a cognitive radio network, we proved the performance gains of combining WEH, IA, and TW relaying. {As the future direction, using full-duplex TW relays instead of the half-duplex TW relays for both PU and SU pairs can be considered to further increase the spectral efficiency and reduce the transmission delay in the system.}

\section*{Appendix}
The optimization function in \eqref{eq:23} is written by
\begin{equation}\label{eq:29}
\begin{split}
&F^{[j]}_{R_S}=E\{\|\mathbf{f}_{\!_{R_S}}^{[j]}-\mathbf{x}_{{i}}\|^2\}= \\
&E\{Tr((\sqrt{p_{{i}}r_{\!_{{R_S}{i}}}^{-\tau}}\mathbf{U}_{R_S}^{[j]\emph{H}}\mathbf{H}_{{R_S}{i}}\mathbf{V}_{i}\mathbf{x}_{i}\\
&+\sqrt{p_{\!_{P_j}}r_{\!_{{R_S}{P_j}}}^{-\tau}}\mathbf{U}_{R_S}^{[j]\emph{H}}\mathbf{H}_{{R_S}{P_j}}\mathbf{V}_{P_j}\mathbf{x}_{\!_{P_j}} \\
&+\mathbf{U}_{R_S}^{[j]\emph{H}}\mathbf{n}^{[j]}_{\!_{R_S}}-\mathbf{x}_{{i}}) \\
&(\sqrt{p_{{i}}r_{\!_{{R_S}{i}}}^{-\tau}}\mathbf{U}_{R_S}^{[j]\emph{H}}\mathbf{H}_{{R_S}{i}}\mathbf{V}_{i}\mathbf{x}_{{i}} \\
&+\sqrt{p_{\!_{P_j}}r_{\!_{{R_S}{P_j}}}^{-\tau}}\mathbf{U}_{R_S}^{[j]\emph{H}}\mathbf{H}_{{R_S}{P_j}}\mathbf{V}_{P_j}\mathbf{x}_{\!_{P_j}}\\
&+\mathbf{U}_{R_S}^{[j]\emph{H}}\mathbf{n}^{[j]}_{\!_{R_S}}-\mathbf{x}_{{i}})^{H})\},
\end{split}
\end{equation}
in which $i=A$ for $j=1$, and $i=B$ for $j=2$. To achieve the IA decoding matrix of the secondary relay at the first two time-slots ($\mathbf{U}^{[j]}_{R_S}$, $j \in \{1,2\}$) the derivative of \eqref{eq:29} with respect to $\mathbf{U}^{[j]}_{R_S}$ should be set equal to zero. It is worth mentioning that because of the i.i.d. symbols, we have $E\{\mathbf{x}_{\!_m}\mathbf{x}_{\!_n}^\emph{H}\}=\mathbf{0}_{d_m \times d_n}$, for $m\neq n$ , and $E\{\mathbf{x}_{\!_m}\mathbf{x}_{\!_m}^\emph{H}\}=\mathbf{I}_{N_m}$ for $m,n \in \{A,B,R_S,P_1,P_2,R_P\}$. Thus, we have
\begin{equation}\label{eq:30}
\begin{split}
&\frac{\delta F^{[j]}_{R_S}}{\delta \mathbf{U}^{[j]}_{R_S}}= \mathbf{U}_{R_S}^{[j]\emph{H}}(p_{{i}}r_{\!_{{R_S}{i}}}^{-\tau}\mathbf{H}_{{R_S}{i}}\mathbf{V}_{i}\mathbf{V}^{\emph{H}}_{i}\mathbf{H}^{\emph{H}}_{{R_S}{i}}\\
&+p_{\!_{P_j}}r_{\!_{{R_S}{P_j}}}^{-\tau}\mathbf{H}_{{R_S}{P_j}}\mathbf{V}_{P_j}{V^{\emph{H}}_{P_j}}\mathbf{H}^{\emph{H}}_{{R_S}{P_j}}+\mathbf{I}_{N_{R_S}})\\
&-\sqrt{p_{{i}}r_{\!_{{R_S}{i}}}^{-\tau}}\mathbf{V}^{\emph{H}}_{i}\mathbf{H}^{\emph{H}}_{{R_S}{i}}=0.
\end{split}
\end{equation}
And in this way, \eqref{eq:25} will be achieved.

\bibliography{Preprint}
\bibliographystyle{ieeetr}

\end{document}